\documentclass[aps,pra,twocolumn,groupedaddress,showpacs, floatfix]{revtex4-1}
\usepackage{amsmath}
\usepackage{graphicx}
\usepackage{bm}
\usepackage[mathlines]{lineno}

\begin{document}

\title{Measuring the Rydberg Constant Using Circular Rydberg Atoms in an Intensity-Modulated Optical Lattice}
\author{Andira Ramos$^1$} \email{andramos@umich.edu} 
\author{Kaitlin Moore$^2$}
\author{Georg Raithel$^{1,2}$}
\affiliation{$^1$Department of Physics, University of Michigan, Ann Arbor, Michigan 48105, USA}
\affiliation{$^2$Applied Physics Program, University of Michigan, Ann Arbor, Michigan 48105, USA}

\date{\today}

\begin{abstract}

A method for performing a precision measurement of the Rydberg constant, $R_{\infty}$, using cold circular Rydberg atoms is proposed. These states have  long lifetimes, as well as negligible quantum-electrodynamics (QED) and no nuclear-overlap corrections. Due to these advantages, the measurement can help solve the ``proton radius puzzle" [Bernauer, Pohl, Sci. Am. 310, 32 (2014)]. The atoms are trapped using a Rydberg-atom optical lattice, and transitions are driven using a recently-demonstrated lattice-modulation technique to perform Doppler-free spectroscopy. The circular-state transition frequency yields $R_{\infty}$. Laser wavelengths and beam geometries are selected such that the lattice-induced transition shift is minimized. The selected transitions have no first-order Zeeman and Stark corrections, leaving only manageable second-order Zeeman and Stark shifts. For Rb, the projected relative uncertainty of $R_{\infty}$ in a measurement under the presence of the Earth's gravity is $10^{-11}$, with the main contribution coming from the residual lattice shift. This could be reduced in a future micro-gravity implementation. The next-important systematic arises from the Rb$^+$ polarizability (relative-uncertainty contribution of $\approx 3 \times10^{-12}$).

\end{abstract}

\maketitle

\section{Introduction}

Knowing the value of the Rydberg constant ($R_{\infty}$) accurately has been of interest for decades due to its relation to other fundamental constants and its role in calculations of atomic energy levels. More recently, the large discrepancies in the proton \cite{Pohl2010,Hill2017} and deuteron \cite{Pohl2016} radii that were found using muonic hydrogen and deuterium, respectively, have reinforced the need to confirm the accuracy of $R_{\infty}$. Previous precision experiments with this goal have involved low-lying states of hydrogen, limited typically by statistical uncertainties, AC Stark shifts and second-order Doppler shifts \cite{CODATA2002}. These have led to the current CODATA relative uncertainty for the Rydberg constant value of $5.9 \times 10^{-12}$ \cite{CODATA2014}. There has also been a study involving circular Rydberg states of hydrogen \cite{DeVries} (relative uncertainty of $2.1 \times 10^{-11}$)  and a proposal involving circular states of lithium \cite{Haroche93} (expected relative uncertainty of about $10^{-10}$). The approaches involving low-lying states and circular states deal with significantly different frequency regimes: optical versus microwave. Therefore, measurements involving low-lying states of hydrogen can have a better relative uncertainty $\delta \nu / \nu$ than results for circular states (under the assumption of similar absolute uncertainty, $\delta \nu$). However, circular states are insensitive to several systematics that are limiting in spectroscopy of low-lying states, as discussed in this paper. It has also been proposed to measure the Rydberg constant using high-angular momentum states of hydrogen-like ions \cite{Jentschura2010}.

\begin{figure}[h!]
\begin{center}
\includegraphics[width=3.2in]{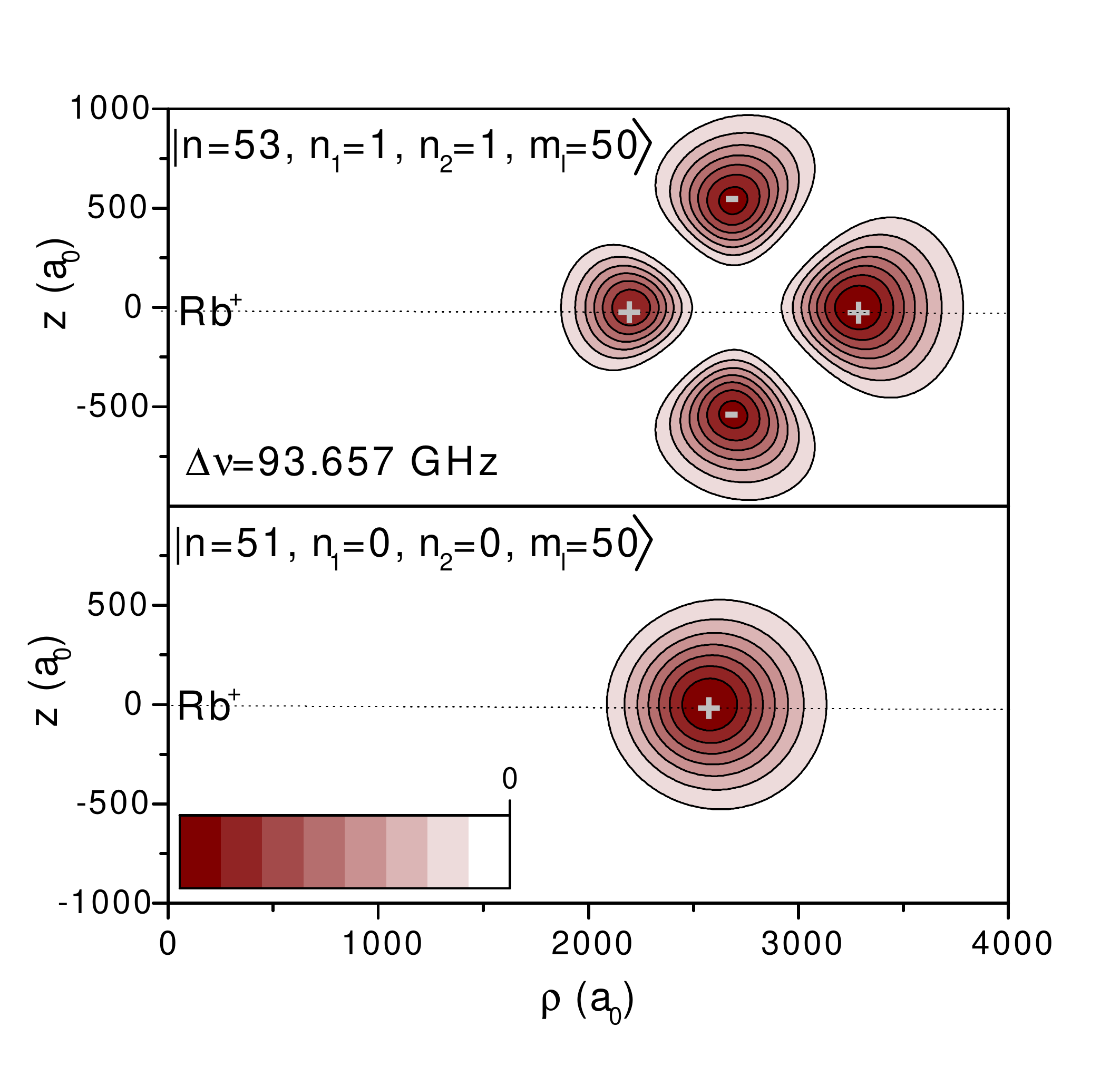}
\caption{Normalized probability density, $|\psi|^2$, for the circular state (bottom) and near-circular state (top) of the transition of interest in the proposed experiment. The kets are labeled in principal, parabolic and magnetic quantum numbers, $n$, $n_{\textrm{1}}$, $n_{\textrm{2}}$, $m$ ~\cite{BetheBook}. Near the Rb$^+$ core, $|\psi|^2 =0$ for both states. The signs refer to the polarity of $\psi$. It is seen that the transition requires a quadrupolar interaction.}
\label{fig:wavefunction}
\vspace{-0.3 in}
\end{center}
\end{figure}

We propose an experiment to obtain an independent measurement of the Rydberg constant using cold, trapped circular Rydberg atoms. Transitions are driven using a recently-developed spectroscopic method, which is based on lattice modulation at microwave frequencies \cite{Moore2015}.  The critical advantages of circular states, in comparison with low-lying states of hydrogen, are their long radiative lifetimes (on the order of ms) \cite{GallagherBook}, their small QED corrections \cite{Wundt2008}, and the absence of any overlap with the nucleus, hence eliminating nuclear charge distribution effects (see Fig.~\ref{fig:wavefunction}).

In contrast with previous efforts to measure $R_{\infty}$ \cite{CODATA2002,CODATA2012}, it is proposed to trap cold Rydberg atoms using a ponderomotive potential optical lattice (POL) \cite{Anderson2011} instead of using cold atomic beams. Trapping the atoms allows for increased interaction times. The light shift introduced by the lattice trap can be addressed by the use of ``magic" conditions under which the trap-induced shifts of upper and lower states cancel \cite{Takamoto2005}. Residual imperfections in the magic-lattice trap can be addressed by performing the experiment under micro-gravity conditions, which allow for an overall reduction in the magic-lattice shift.

A key method is ponderomotive spectroscopy \cite{Moore2015}, which allows for circular-to-near-circular transitions to be driven by amplitude-modulating the ponderomotive optical lattice. The method permits us to drive the quadrupolar transition needed in our work as a first-order process and to eliminate Doppler broadening. By driving transitions between states of the same magnetic quantum number, we eliminate the first-order Zeeman effect. To obtain a zero first-order Stark shift, we select states with parabolic quantum numbers $n_{\textrm{1}}=n_{\textrm{2}}$.

In these ways, the proposed experiment addresses important issues that have been encountered in preceding Rydberg constant measurements. In addition, due to the elimination of nuclear charge overlap, the measurement of $R_{\infty}$ is also independent of the radius of the proton. Overall, a relative uncertainty of $10^{-11}$ is expected, which would already be enough to shed light onto the proton radius puzzle. An improvement beyond the current CODATA uncertainty, $\delta \nu / \nu =5.9 \times 10^{-12}$~\cite{CODATA2014}, is possible by an implementation of the experiment under micro-gravity conditions.

\section{Proposed Experimental Outline}

\subsection{Atom Preparation and Spectroscopy}

We use cold atoms to reduce interaction-time broadening and to limit the interaction volume, thereby reducing field inhomogeneity effects. We propose to use $^{85}$Rb atoms, pre-cool them in a MOT to $\sim$ 100 $\mu$K, and further cool them to  $\sim$ 10 $\mu$K and  $\sim$ 1 $\mu$K in bright and gray optical molasses, respectively  \cite{Salomon1990, Salomon1996}. Since the atom densities desired for this experiment are moderate ($\lesssim 10^8$ cm$^{-3}$), to avoid Rydberg-Rydberg interactions, the use of optical molasses is ideal.

\begin{figure}[t]
\includegraphics[width=3.4in]{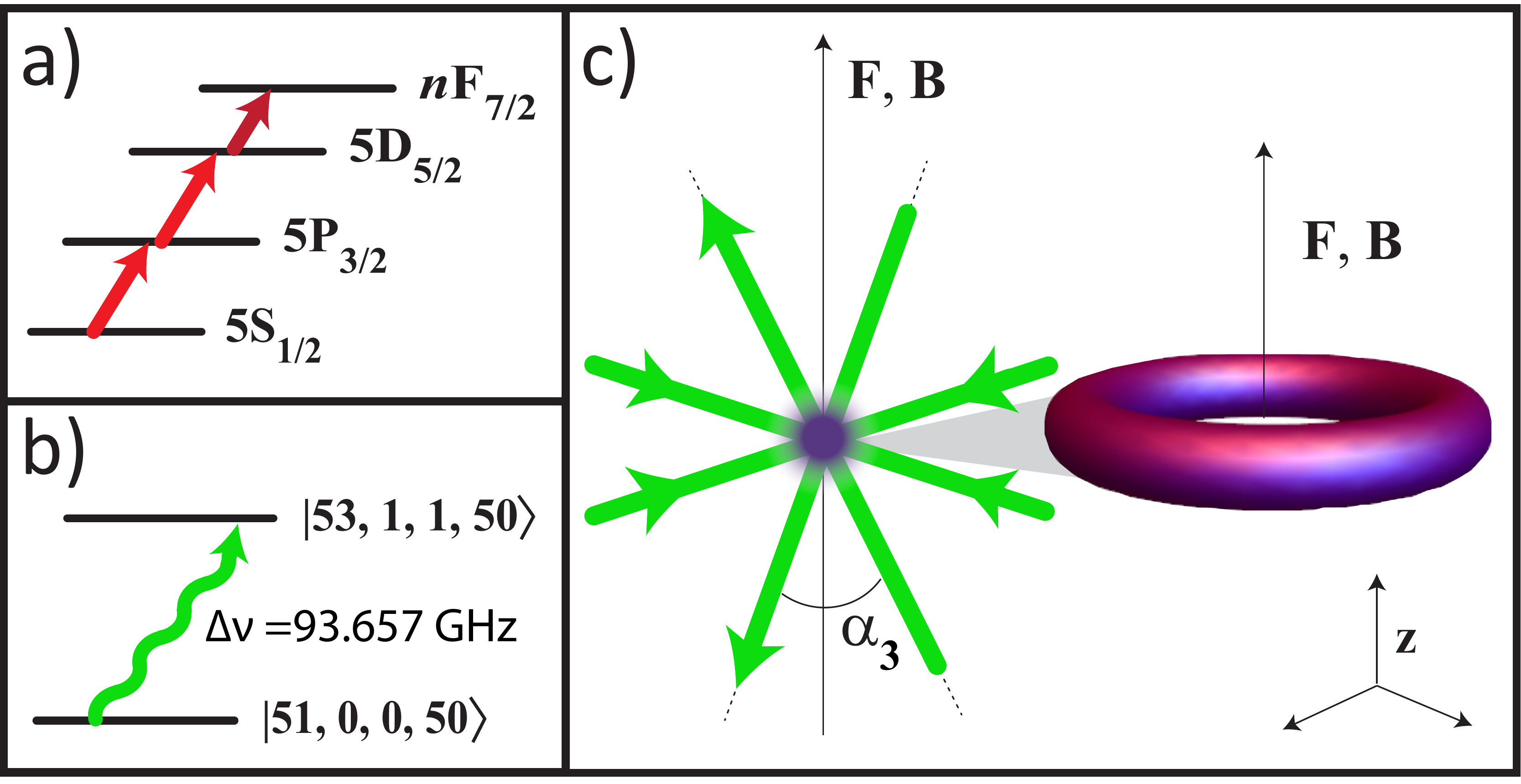}
\caption{(color online). a) Excitation scheme used to prepare atoms before circularization. b) After circularization, the transition between an initial, circular state and a final, near-circular state (expressed using parabolic quantum numbers $|n, n_{\textrm{1}}, n_{\textrm{2}}, m_{\textrm{l}}\rangle$) is driven by ponderomotive spectroscopy, indicated by the curly arrow. c) Parallel stabilization fields, $\bf{F}$ and $\bf{B}$, define the quantization axis ($z$-axis) of the atoms.  The angles between different pairs of counter-propagating lattice beams, $\alpha_{\textrm{i}}$, can be varied to change the periodicities of the lattice in the $i^{th}$ direction.}
\label{fig:setup}
\vspace{-0.1 in}
\end{figure}

In order to circularize atoms, we use a modified adiabatic rapid passage (ARP) method \cite{ARPHaroche, ARPKleppner}, which in Rb requires that the Rydberg atoms are initially prepared in $m_l$=3. We employ a three-level excitation scheme, $5S_{1/2} \rightarrow 5P_{3/2}$ (wavelength of 780~nm), $5P_{3/2} \rightarrow 5D_{5/2}$ (776~nm), and $5D_{5/2} \rightarrow nF_{7/2}$ (1260~nm) (see Fig.~\ref{fig:setup}a). The modified ARP circularization method is optimal for states $n\lesssim50$ (our case). If much higher principal quantum numbers were desired, the ``\textbf{E}$\times$\textbf{B}" method \cite{Goy1988,Delande1988} could be used.

In order to take advantage of the circular atoms' long lifetimes (tens of ms at 4 K), it is necessary to trap the Rydberg atoms. To achieve this, a three-dimensional standing-wave optical lattice is adiabatically turned on, and the atoms are trapped via the ponderomotive potential \cite{Dutta2000} (see Fig.~\ref{fig:setup}c). This ponderomotive potential emerges from the last term in the minimal-coupling Hamiltonian (in SI units),
\begin{equation}
\label{eq:MinHamiltonian}
\hat{H}=\frac{1}{2m_{\textrm{e}}}(2|e|\mathbf{A}(\mathbf{\hat{r}})\cdot \mathbf{\hat{p}}+ e^2\mathbf{A}(\mathbf{\hat{r}})\cdot \mathbf{A}(\mathbf{\hat{r}})),
\end{equation}
\begin{figure}[b]
\begin{center}
\includegraphics[width=3.4in]{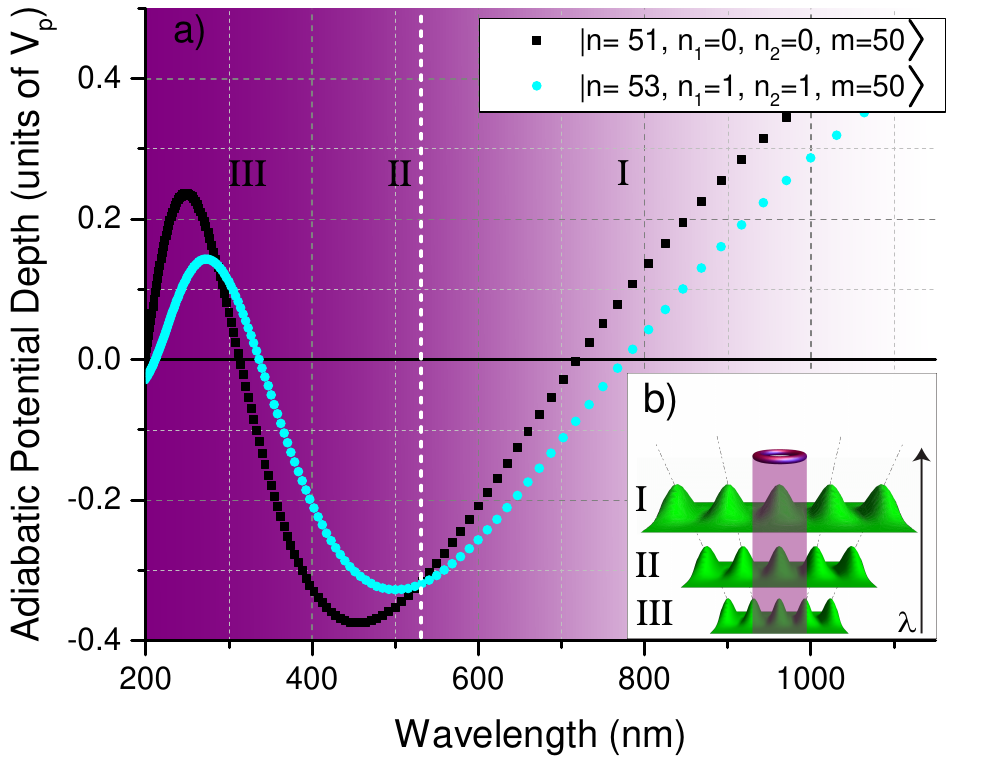}
\caption{a) Ponderomotive adiabatic potential depth (in units of the free electron potential, $V_{\textrm{p}}$, the energy of a free electron in the lattice laser field) as a function of wavelength for the two states of interest in a one-dimensional lattice formed by counter-propagating beams ($\alpha =0$). The points at which the two plots cross are the ``magic" wavelengths for this particular pair of states. The magic wavelength we choose for this experiment is shown with a white dashed line, and it occurs at about 532~nm. b) Schematics of the projection of the wavefunction density onto the lattice, as the wavelength is varied and the atom size remains fixed. Labels I, II and III correspond to those in part a). The oscillatory behavior and flip in signs in a) are related to how many lattice periods fall within the volume of the atom \cite{Anderson2011}.}
\label{fig:magic}
\end{center}
\end{figure}

\noindent which is proportional to laser intensity and arises when a quasi-free Rydberg electron is placed in a rapidly oscillating field. Here, $\mathbf{A}(\mathbf{\hat{r}})$ is the vector potential of the optical field, $\mathbf{\hat{p}}$ is the momentum operator, $\mathbf{\hat{r}}$ is the position operator, $m_{\textrm{e}}$ is the electron mass and $e$ is the electron charge. For non-degenerate states in a monochromatic lattice, the adiabatic trapping potential is given by 
\begin{equation}
\label{eq:AdotA}
V_{\textrm{ad}}(\mathbf{R})= \frac{e^2}{2m_{\textrm{e}}}\int|\psi(\mathbf{r})|^2 |\mathbf{A}(\mathbf{r}+\mathbf{R})|^2 d^3r,
\end{equation}
\noindent where $\mathbf{R}$ is the atomic center-of-mass position.

The wavelength of the lattice is chosen to match a ``magic" condition for the desired transition (equal potentials for lower and upper states). To illustrate this, in Fig.~\ref{fig:magic}a we show adiabatic lattice-potential depths obtained from Eq. ~\ref{eq:AdotA} for the two states of interest, $\textit{n}$=51 and $\textit{n}$=53, as a function of wavelength. In the figure it is seen that for the two states considered there are two options to achieve magic conditions, 290~nm and 532~nm. For the proposed experiment we choose a 532-nm lattice.

The same ponderomotive term that traps atoms is also used to drive transitions between circular and near-circular states \cite{Moore2015}. In ponderomotive spectroscopy, the optical-field intensity varies substantially within the volume of the atom, and the lattice amplitude modulation frequency is resonant with an atomic transition or a sub-harmonic of it \cite{MooreMagic}. In the proposed experiment, the spatial variation of the lattice is about the same as the size of the atom, and the lattice-amplitude modulation frequency (atomic transition frequency) is about 100 GHz. Finally, the population transfer between initial and final states is measured using state-selective field ionization~\cite{GallagherBook}.

The proposed experiment occurs inside a cryogenic enclosure that provides a radiation temperature near 4~K. This is done to decrease blackbody radiation effects, which reduce the Rydberg-state lifetimes~\cite{GallagherBook} but cause only minor shifts of the transition frequencies of interest (see section~III.G).

\subsection{Stabilization Fields}

The $\lesssim$2$n^2$-fold degeneracy of the circular and hydrogenic states must be lifted using stabilization fields. To stabilize the circular Rydberg states against state-mixing, we employ an electric field, $F$, and a weak, parallel magnetic field, $B$ ($B/2<(3/2)nF<n^{-3}$, in atomic units). This stabilization method is suitable for high-precision spectroscopy \cite{Dutta2000, DeVries}.

In order to probe the atoms, we trap them in a ponderomotive potential. The ponderomotive shift must be smaller than the shift caused by the stabilizing electric and magnetic fields \cite{Dutta2000}. The potential must also be deep enough to trap atoms at their thermal kinetic energy. Hence, the laser-cooling temperature sets minimum values for the fields that we use in both trapping and stabilization of our states. The hierarchy of shifts is shown in Table~\ref{tab:shiftScales} for MOT temperatures ($\sim$ 100$\mu$K), temperatures in gray optical molasses ($\sim$ 1$\mu$K)  \cite{Salomon1990,Salomon1996} and Bose-Einstein condensate (BEC) temperatures ($\sim$ 10 ~nK). Table~\ref{tab:fields} shows the corresponding typical field magnitudes and provides guidance in designing the circular-state stabilization scheme.

\begin{table}[h!]
\caption{\label{tab:shiftScales} Hierarchy of level shifts in three atomic temperature regimes.\footnote{All energies are expressed in kHz.}}
\begin{tabular}{ lllll}
T  ($\mu$K) & Thermal Energy \footnote{$\textrm{Thermal Energy}=k_{\textrm{B}}T/2$.}  & POL  & Magnetic  & Electric \\
\hline
100  		 & 1000 		   & 3100 	      & 9400 		 & 28000 \\
1      		 & 10    		   & 31 	      & 94 		 & 280 \\
0.01          	 & 0.1   		   & 0.31	      & 0.94        	 & 2.8 \\
\end{tabular}
\end{table}

\begin{table}[h!]
\caption{\label{tab:fields} Magnetic and electric fields suitable for three temperature regimes. The fields satisfy optical-trap depth $\approx k_\textrm{B} T<B/2<(3/2)nF$, for $n$=51.}
\begin{tabular}{ lll }
T ($\mu$K)& Magnetic Field (mT)  & Electric Field (mV/cm) \\
\hline
100  	  & 0.67           & 290 \\
1       	  & $6.7 \times 10^{-3} $      & 2.9 \\
0.01     & $6.7 \times 10^{-5}$    & 0.029\\
\end{tabular}
\end{table}

Since we are always in the Paschen-Back regime of the fine structure (see sections~III~B and~E), we use the basis set $\{|n, n_{\textrm{1}}, n_{\textrm{2}},m_{\textrm{l}}, m_{\textrm{s}}\rangle\}$ \cite{BetheBook}  throughout this paper, where $n_{\textrm{1}}$ and $n_{\textrm{2}}$ are parabolic quantum numbers.

Parabolic and spherical bases are related by \cite{GallagherBook}
\begin{equation}
\label{eq:states}
 |n, n_{\textrm{1}}, n_{\textrm{2}}, m_{\textrm{l}} \rangle= \sum_{l} C^{n_{\textrm{1}},n_{\textrm{2}}}_{l,m_{\textrm{l}}} |n,l, m_{\textrm{l}} \rangle,
\end{equation}
\noindent where the Clebsch-Gordan coefficients are related to the Wigner 3J symbols by \cite{GallagherBook} $C^{n_{\textrm{1}} n_{\textrm{2}}}_{l m_{\textrm{l}}}=\langle n, n_{\textrm{1}}, n_{\textrm{2}}, m_{\textrm{l}}|n, l, m_{\textrm{l}} \rangle$,
\begin{equation}
\begin{split}
C^{n_{\textrm{1}}n_{\textrm{2}}}_{lm_{\textrm{l}}} &= (-1)^{(1-n+m_{\textrm{l}}+n_{\textrm{1}}-n_{\textrm{2}})/2+l} \sqrt{2l+1} \\
&\quad \times \begin{pmatrix}
\frac{n-1}{2} & \frac{n-1}{2} & l\\  
\frac{m_{\textrm{l}}+n_{\textrm{1}}-n_{\textrm{2}}}{2} &  \frac{m_{\textrm{l}}-n_{\textrm{1}}+n_{\textrm{2}}}{2} & -m_{\textrm{l}}
\end{pmatrix}.
\end{split}
\end{equation}
Note that for the states in Figs.~\ref{fig:wavefunction},~\ref{fig:setup} and~\ref{fig:magic}, the sum over $l$ has at most two non-zero terms.

\section{Energy Shifts}

In this section we discuss the various energy-level shifts; the results are summarized in the discussion (section~IV).

\subsection{Lattice-Induced Shift}

In its interaction with the optical-lattice field, the Rydberg electron behaves as a quasi-free particle. For a plane-wave linearly-polarized field of the form $\hat{\mathbf{x}}F_L(\mathbf{r}) \sin( \omega t)$ ($\hat{\mathbf{x}}$ is a unit vector), Eq.~\ref{eq:MinHamiltonian} leads to the free-electron ponderomotive potential
\begin{equation}
V_{\mathrm{\textrm{p}}}= \frac{e^2|\mathbf{F}_L (\mathbf{r})|^2}{4 m_{\textrm{e}} \omega^2},
\end{equation}
\noindent where $\omega$ is the angular frequency of the laser electric field and $|\mathbf{F}_L (\mathbf{r})|^2$ is proportional to the spatially-varying field intensity ~\cite{FriedrichBook}. The ponderomotive potential is the average kinetic energy of the free electron in the lattice laser field, and it is polarization- and phase-independent.

The position-dependent ponderomotive potential, $V_{\textrm{p}}(\hat{\mathbf{r}}+\mathbf{R})$, is added as a perturbation to the Rydberg electron's Hamiltonian. Diagonalization of the Rydberg Hamiltonian yields the Born-Oppenheimer (BO) adiabatic potential surfaces, $V_{\textrm{p}}(\mathbf{R})$, for the atom's center-of-mass motion, as well as the associated adiabatic Rydberg-electron wavefunction, $\psi (\mathbf{r};\mathbf{R})$ \cite{Dutta2000}.

Generally, $\psi(\mathbf{r}; \mathbf{R})$ is unknown and must be simultaneously solved for along with the BO potentials~\cite{YoungeNJP}. In our regime, where the shifts due to the parallel electric and magnetic stabilization fields dominate the optical shifts, the adiabatic states are given by the parabolic basis states, $|n, n_{\textrm{1}}, n_{\textrm{2}},m_{\textrm{l}}, m_{\textrm{s}}\rangle$. This greatly simplifies the calculation of the BO adiabatic potential because $\psi(\mathbf{r}; \mathbf{R})$ is no longer dependent on $\mathbf{R}$. Since in our case, the optical lattice is formed by three sets of lattice beams (see Fig.~\ref{fig:setup}), the three-dimensional BO adiabatic potential follows from
\begin{equation}
\label{eq:3DPOL}
\begin{split}
V_{\mathrm{ad}}(\mathbf{R}) &= \int d^3r \sum_i\frac{e^2|F_{L\textrm{i}} \cos(\Delta\mathbf{k}_{\textrm{i}}\cdot(\mathbf{R}+\mathbf{r}))|^2}{m_{\textrm{e}}\omega_{\textrm{i}}^2}\\
&\quad \times | \psi_{n,n_1,n_2}(\mathbf{r})|^2.
\end{split}
\end{equation}
\begin{figure}[t]
\begin{center}
\includegraphics[width=3.4in]{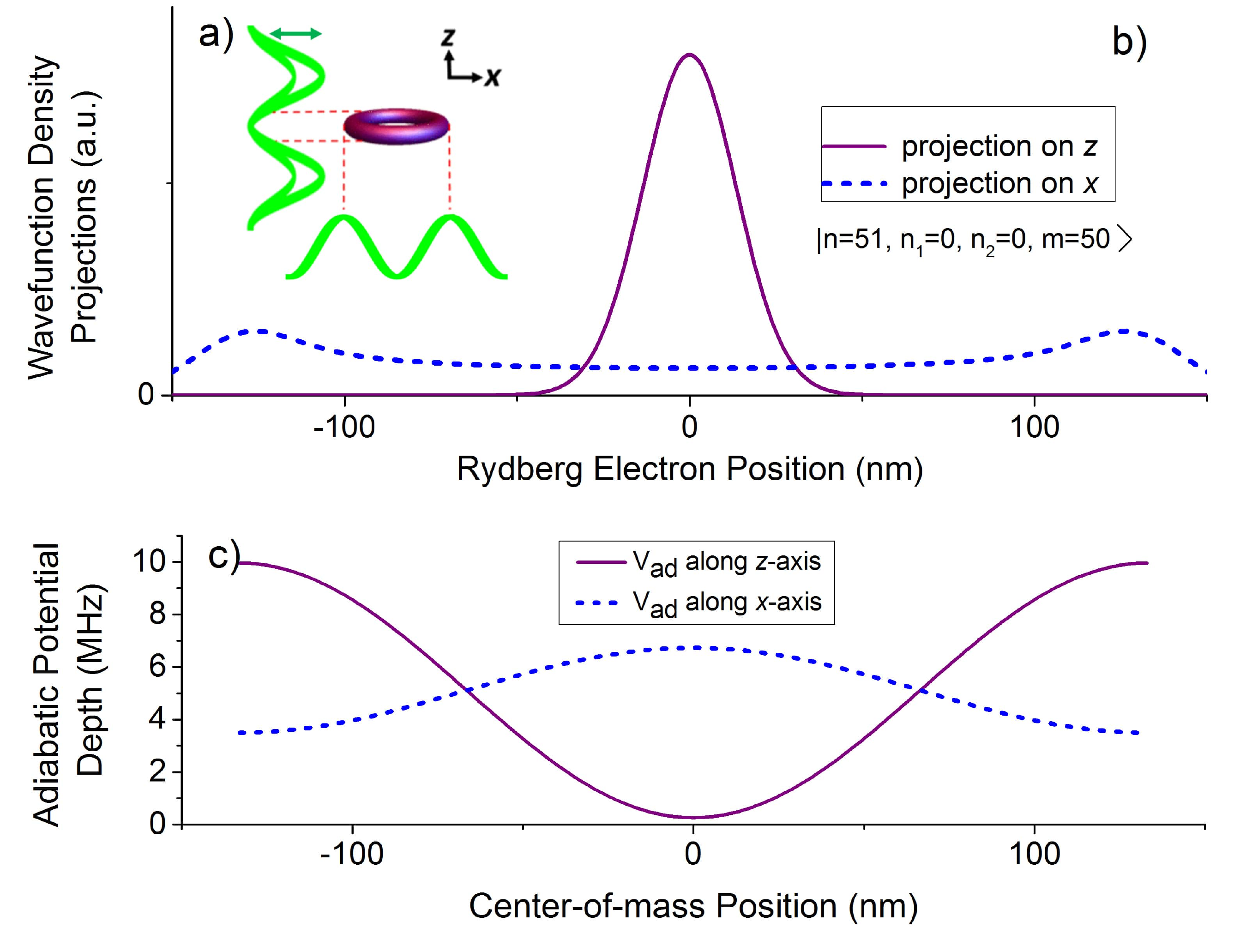}
\caption{\label{fig:potential} Effects of wavefunction projections on the depth of the BO adiabatic potential for a 532-nm lattice extending along $x$ and $z$, with single-beam intensities $4 \times 10^{9}$~W/$\mathrm{m^2}$ ($\alpha_i=0$). a) Alignment of the optical-lattice standing-waves and the circular-state probability distribution. The amplitude of the $z$-direction lattice is modulated in time. b) Projections of $|\psi|^2$ along $x$ and $z$. The overlap of the projections with the optical-lattice standing waves determines the BO adiabatic trapping potentials along the respective coordinate directions (see Eq.~\ref{eq:3DPOL}). c) Trapping potentials (as a function of the center-of-mass position of the atom) calculated from Eq.~\ref{eq:3DPOL}; the zero position corresponds to a lattice field node. The different depths and phases are a result of the quite distinct wavefunction projections onto $x$ and $z$.}
\vspace{-0.2 in}
\end{center}
\end{figure}

In the integral in Eq.~\ref{eq:3DPOL}, $|\psi_{n,n_1,n_2}(\mathbf{r})|^2$ acts as a weighting factor. There, $i$ is the summing index over optical-lattice directions  (for a 3D lattice, $i=1,2,3$), which need not be orthogonal to each other; $F_{Li}$ is the field amplitude of a single beam; $\psi_{n,n_1,n_2}(\mathbf{r})$ is the $\mathbf{R}$-independent Rydberg electron wavefunction; $\mathbf{r}$ is the valence electron (relative) position, $\omega_{\textrm{i}}$ is the angular frequency of the lattice beam; and $|\Delta \mathbf{k}_{\textrm{i}}|=|\mathbf{k}_{\textrm{i1}}-\mathbf{k}_{\textrm{i2}}|= 2k_{\textrm{i}} $cos$(\alpha_{\textrm{i}}/2)$, where $\mathbf{k}_{\textrm{i1}}$ and $\mathbf{k}_{\textrm{i2}}$ are the wavevectors corresponding to the pair of lattice beams along the $i^{th}$ direction, and $\alpha_{\textrm{i}}$ is the angle between a pair of counter-propagating beams (see Fig.~\ref{fig:setup}c).

The ratio of laser intensities of the lattice axes (up to three) and the aspect ratio between the atom's size (defined by its known state) and the optical lattice periodicities (defined by $\lambda_{\textrm{i}}$ and $\alpha_{\textrm{i}}$) (Fig.~\ref{fig:magic} and Fig.~\ref{fig:potential}) can be experimentally controlled. This allows us to vary the depth and the minimum potential value of $V_{\textrm{ad}}(\mathbf{R})$ (see Fig.~\ref{fig:magic}) and to realize a ``magic"-lattice condition (where the two states in the transition experience the same energy shift in the BO potential $V_{\textrm{ad}}(\mathbf{R})$).
	
Experimental considerations suggest to use $\alpha_\textrm{i}=0$ and to choose common laser wavelengths $\lambda_\textrm{i}$, leaving the intensity ratios and the atomic quantum numbers to attain a magic condition. Here we consider the $n=51 \rightarrow n=53$ transition, which, as seen in Fig.~\ref{fig:magic}, has a magic wavelength at about 532~nm (the second harmonic of a ND:YAG laser). Using a magic lattice, most of the lattice-induced shift in the proposed experiment can be eliminated (see section~IV.A).

\subsection{First-Order Zeeman and Stark Shifts}

To avoid undesired state-mixing due to minute stray electric and magnetic fields, stabilization fields must be applied to the atoms. The most suitable stabilization scheme is one where $(3/2)nF>B/2$, in atomic units (see section~II.B). 
The Stark Hamiltonian for an electric field, $\mathbf{F}$, pointing along $z$ is
\begin{equation}
\hat{H}_{\mathrm{S}}= Fe\hat{z},
\end{equation}
\noindent where $\hat{z}$ is the $z$-component of the position operator. In the basis of parabolic states,  ${|n, n_{\textrm{1}}, n_{\textrm{2}}, m_{\textrm{l}}\rangle}$, with quantization axis along $z$, the first-order eigenvalues of the Stark Hamiltonian are
\begin{equation}
E_{\mathrm{S}}= \frac{3}{2} F e a_0 n(n_{\textrm{1}}-n_{\textrm{2}}).
\end{equation}

The states and transitions in this work are of the type $|n, 0,0,n-1\rangle\leftrightarrow|n+2,1,1,n-1\rangle$; in this case the linear Stark shifts for both levels, as well as for the transition, are identical zero.

We apply a weak magnetic field, $B$, in the $z$-direction that removes the remaining degeneracies between states of interest and other Stark levels. This results in a Zeeman Hamiltonian
 \begin{equation}
\hat{H}_{\mathrm{Z}}= \frac{B e}{2 m_{\textrm{e}}} (g_{\textrm{L}} \hat{L}_{\textrm{z}}+g_{\textrm{e}} \hat{S}_{\textrm{z}}),
\end{equation}
\noindent where $g_{\textrm{e}}$ is the electron spin $g$-factor, $g_{\textrm{L}}$ is the electron's orbital $g$-factor ($g_{\textrm{L}}=1$) and $\hat{L}_{\textrm{z}}$ and $\hat{S}_{\textrm{z}}$ are the orbital angular momentum and spin angular momentum operators, respectively. In the Paschen-Back regime, the parabolic states with spin,  ${|n, n_{\textrm{1}}, n_{\textrm{2}}, m_{\textrm{l}}, m_{\textrm{s}}\rangle}$, are eigenstates of the Zeeman Hamiltonian. This shift is given by
\begin{equation}
E_{\mathrm{Z}} = \frac{B \hbar e}{2 m_{\textrm{e}}} (m_{\textrm{l}}+ g_{\textrm{e}} m_{\textrm{s}}),
\end{equation}
\noindent where $\hbar$ is the reduced Planck's constant. Since $m_{\textrm{l}}$ and $m_{\textrm{s}}$ are equal for both states involved in the transition considered, the linear Zeeman shift of the transition is zero.

It is critical that the angle between the electric and magnetic field be close to zero for this to hold, since any departure from zero would introduce $x$- or $y$-components of the fields, leading to the appearance of additional second-order shifts. These can be estimated using Eq. 2.15 in \cite{DeVries}, which yields an upper limit for the allowed angular misalignment between the fields of about one degree.


\subsection{Second-Order Stark and Diamagnetic Shifts}

Second-order perturbation theory for the Stark effect Hamiltonian yields an energy shift of \cite{GallagherBook,DeVries}
\begin{equation}
\begin{split}
E_{\mathrm{SS}} &= \frac{-4 \pi\epsilon_{\textrm{0}} a_0^3 F^2n^4}{16} \\
&\quad \times \left [ 17n^2-3(n_{\textrm{1}}-n_{\textrm{2}})^2 -9m_{\textrm{l}}^2+19 \right ] ,
\end{split}
\end{equation}
\noindent which is small (see Table~\ref{tab:budget}). The diamagnetic Hamiltonian is
 \begin{equation}
\hat{H}_{\mathrm{D}}= \frac{e^2 B^2}{8m_{\textrm{e}}} (\hat{x}^2 +\hat{y}^2),
\end{equation}
\noindent where $\hat{x}$ and $\hat{y}$ are the $x$ and $y$-direction position operators. This Hamiltonian can be rewritten in the spherical basis as

\begin{equation}
\label{eq:diagZeeman}
\hat{H}_{\mathrm{D}}= \frac{e^2 B^2}{8 m_{\textrm{e}}} (\hat{r}^2 \sin^2\hat{\theta}),
\end{equation}

\noindent where the operator $\hat{\theta}$ is the angle with respect to the $z$-axis. Using Eq.~\ref{eq:diagZeeman}, we obtain a diamagnetic energy shift of
\begin{equation}
E_{\mathrm{SZ}} = \sum_{l}\frac{e^2 B^2}{8 m_{\textrm{e}}} |C^{n_{\textrm{1}}n_{\textrm{2}}}_{lm_{\textrm{l}}}|^2  \left \langle n l m_{\textrm{l}}|\hat{r}^2 \sin^2\hat{\theta} | n l m_{\textrm{l}} \right \rangle,
\end{equation}
\noindent where the angular matrix elements are given in \cite{BetheBook} and the radial matrix elements can be computed numerically. As shown in Table ~\ref{tab:budget}, these second-order shifts lead to uncertainty contributions below the current uncertainty goal.

\subsection{Quantum Defects}

In alkali atoms, polarization and penetration quantum defects are introduced as corrections to the hydrogenic eigenvalue \cite{Kleppner1976, GallagherBook}
\begin{equation}
E=-hcR_{\textrm{Rb}}\frac{1}{(n-\delta _{\textrm{l}})^2},
\end{equation}
\noindent where $c$ is the speed of light, $n$ is the principal quantum number, $R_{\textrm{Rb}}= \frac{M}{m_{\textrm{e}}+M}$~$R_{\infty}$ ($M$ is the mass of Rb$^+$) and $\delta_{\textrm{l}}$ is the quantum defect. This $\delta_{\textrm{l}}$ can be expressed as the sum of the polarization and penetration quantum defects, $\delta_{\textrm{l}} = \delta_{\mathrm{pol}}+ \delta_{\mathrm{pen}}$, which is commonly expanded by using the Rydberg-Ritz formula \cite{GallagherBook}.

This $\delta_{\textrm{l}}$ decreases significantly with increasing $\textit{l}$. In the proposed experiment, where transitions from circular to near-circular states are driven, $\delta_{\textrm{pen}}=0$ because the probability density of circular Rydberg states is zero in the ionic core region (Fig.~\ref{fig:wavefunction}). Core polarization, however, must still be considered, with the shift due to the effective dipole polarizability, $\alpha'_{\textrm{d}}$, being the leading term, followed by an almost negligible shift due to the effective quadrupole polarizability, $\alpha'_{\textrm{q}}$. The effective polarizabilities consist of the DC polarizability and a non-adiabatic correction. The polarization potential is given by \cite{Curtis2003}
\begin{equation}
\label{eq:Vpol}
\hat{V}_{\mathrm{pol}}= \frac{-e^2}{16\pi^2 \epsilon_0^2} \left [ \frac{1}{2}\frac{\alpha'_{\textrm{d}}}{\hat{r}^4}+\frac{1}{2}\frac{\alpha'_{\textrm{q}}}{\hat{r}^6} +  ...  \right ],
\end{equation}
\noindent where the values of $\alpha'_{\textrm{d}}$ and  $\alpha'_{\textrm{q}}$ are obtained from \cite{GallagherNew} and can be converted to SI units \cite{Safronova2010}. For $l \ll n$, $\delta_{\textrm{pol}}$ corresponding to this potential is approximately (in atomic units)
\begin{equation}
\begin{split}
\delta_{\mathrm{pol}} &\approx  \frac{3}{4}\frac{\alpha'_{\textrm{d}}}{l^5}\left [ 1- O_{\textrm{d}}\left ( \frac{l^2}{n^2} \right ) \right ] \\
&\quad  +\frac{35}{16}\frac{\alpha'_{\textrm{q}}}{l^9}\left [ 1- O_{\textrm{q}}\left ( \frac{l^2}{n^2} \right ) \right ],
\end{split}
\end{equation}
\noindent where the $O(l^2/n^2)$ terms are corrections that can be exactly resolved by using the analytically known expressions for $\left \langle r^{-4}\right \rangle$ and $\left \langle r^{-6}\right \rangle$ (see appendix of reference \cite{Kleppner1976}). Since in the proposed experiment high-angular-momentum states are employed, the exact analytic expressions for  $\left \langle r^{-4}\right \rangle$ and $\left \langle r^{-6}\right \rangle$ need to be used.

The quadrupole term in Eq.~\ref{eq:Vpol} becomes negligible at large $\textit{r}$ values such as the ones found in circular Rydberg states. The dipole polarizability term needs to be carefully considered as it leads to corrections of the order of several kHz.

\subsubsection{Polarizabilities}

The polarizabilities in the polarization quantum defect are not well known. The most recent experimental limits are $\alpha'_{\textrm{d}}=9.12$  and $\alpha'_{\textrm{q}}=14$ (in atomic units) \cite{GallagherNew}, which are not consistent with previous theory work \cite{Heinrichs1970, Sternheimer1970} for the dipole and quadrupole polarizabilities, respectively. The current uncertainties in the experimental values of polarizabilities are of order $10^{-3}$, which lead to a relative uncertainty of the order of $10^{-12}$ in the proposed Rydberg constant measurement, making this one of the main sources of uncertainty.

\subsubsection{Non-adiabatic Effects}

The quantum defect theory discussed so far assumes that the Rb$^+$ response to the Rydberg electron's field is adiabatic. However, this is not necessarily the case. The non-adiabaticity of the electron's motion makes it necessary to redefine $\hat{V}_{\mathrm{pol}}$ \cite{Opik1967} and hence the polarizabilities as
\begin{equation}
\hat{V}_{\mathrm{pol}} = \frac{-e^2}{16\pi^2 \epsilon_0^2} \left [ \frac{1}{2}\frac{\alpha_{\textrm{d}} \ y^d_0}{\hat{r}^4}+
\frac{1}{2}\frac{\alpha_{\textrm{q}} \ y^q_0 + \alpha_{\textrm{d}} \ y^d_1}{\hat{r}^6} + ... \right ],
\end{equation}
\noindent where $y^d_0$, $y^d_1$, and $y^q_0$ vary slowly with $\textit{n}$ and $\textit{l}$. Comparing this expression to Eq.~\ref{eq:Vpol}, we see that the corrected and the adiabatic polarizabilities are related as follows: $\alpha'_{\textrm{d}}= y_0^d \alpha_{\textrm{d}} ,\alpha'_{\textrm{q}}=y_0^q \alpha_{\textrm{q}}+y_1^d \alpha_{\textrm{d}}$ \cite{Kleppner1976}. Using Ref.~\cite{Patil1985} the corrected dipolar and quadrupolar polarizabilities can be calculated for $^{85}$Rb. The experimental values of the polarizability used for this work already include the non-adiabatic correction.

\subsection{Fine Structure Correction}

For Rydberg atoms with large $l$, the form of the fine-structure shift is the same as for the hydrogen atom. It has two contributions, the relativistic mass correction and the spin-orbit coupling \cite{SakuraiBook}. The spin-orbit Hamiltonian is
\begin{equation}
\label{eq:FShamiltonian}
\hat{H}_{\mathrm{SO}}= \frac{\alpha \hbar }{2 m_{\textrm{e}}^2c}\frac{1}{\hat{r}^3} \hat{\mathbf{L}}\cdot \hat{\mathbf{S}},
\end{equation}
\noindent where $\alpha$ is the fine-structure constant. In the Paschen-Back regime (our case) the fine-structure-induced correction to the energy levels is
\begin{equation}
\label{eq:FSshift}
E_{\mathrm{SO}} = \frac{\alpha^4 m_{\textrm{e}} c^2}{2} \sum_{l}  | C^{n_{\textrm{1}}n_{\textrm{2}}}_{lm_{\textrm{l}}} |^2 \frac{m_{\textrm{l}} m_{\textrm{s}}}{n^3 l (l+1)(l+\frac{1}{2})}.
\end{equation}

The relativistic contribution to the fine-structure shift results from expanding the expression for the relativistic kinetic energy of a particle. This yields a correction Hamiltonian of ~\cite{SakuraiBook}
\begin{equation}
\hat{H}_{\mathrm{rel}}= - \frac{\hat{\mathbf{p}}^4}{8m_{\textrm{e}}^3 c^2}.
\end{equation}

Following the same procedure presented in \cite{SakuraiBook}, in first-order perturbation theory the relativistic shift is
\begin{equation}
\begin{split}
E_{\mathrm{rel}} &= -\frac{\alpha^4 m_{\textrm{e}} c^2}{2} \sum_{l} | C^{n_{\textrm{1}}n_{\textrm{2}}}_{lm_{\textrm{l}}} |^2 \\
&\quad \times \left [ \frac{1}{n^3(l+\frac{1}{2})}-\frac{3}{4n^4}\right].
\end{split}
\end{equation}

Putting both terms together we obtain the fine-structure energy shift
 \begin{equation}
\begin{split}
E_{\mathrm{FS}} &= -\frac{\alpha^4 m_{\textrm{e}} c^2}{2 n^3} \sum_{l} | C^{n_{\textrm{1}}n_{\textrm{2}}}_{lm_{\textrm{l}}} |^2  \\
&\quad \times \left [ \frac{-m_{\textrm{l}} m_{\textrm{s}}}{l(l+1)(l+\frac{1}{2})}  + \left ( \frac{1}{(l+\frac{1}{2})}-\frac{3}{4n}\right)  \right ].
\end{split}
\end{equation}

For our states of interest, the relativistic correction is around 6~kHz while the spin-orbit correction is around 200~Hz.

\subsection{Quantum Electrodynamic Corrections}

Quantum electrodynamics (QED) introduces the self-energy and the vacuum polarization QED corrections, which together form the Lamb shift. For circular Rydberg states, a first-order account of QED corrections is sufficient; the result is
\begin{equation}
\begin{split}
\label{eq:QED}
E_\mathrm{Lamb} &= \frac{8Z^4\alpha^3}{3\pi n^3} hcR_{\infty} \sum_{l} |C^{n_{\textrm{1}}n_{\textrm{2}}}_{lm_{\textrm{l}}}|^2 \\
&\quad \times \left [ L(n,l) + \frac{3}{8}\frac{c_{\textrm{lj}}}{2l+1} \right ],
\end{split}
\end{equation}
\noindent where
\begin{equation}
c_{\textrm{lj}}=\left\{\begin{matrix}
(l+1)^{-1} \ \mathrm{for} \ j= l+ 1/2\\ 
-l^{-1} \ \mathrm{for} \ j= l- 1/2\\ 
\end{matrix}\right.
\end{equation}
The Bethe logarithm is $L(n,l)$ \cite{Erickson1977}, which can be extrapolated for $n\geq 4, l\geq 3$ as
\begin{equation}
\begin{split}
L(n,l) &= \frac{0.1623834}{2l+1}\left [ \left ( \frac{1}{l} \right )^{3/2} - \left ( \frac{1}{n} \right )^{3/2}\right ] \\
&\quad \times \left [ 1\pm \left ( \frac{1}{2}-\frac{1}{4}\left ( \frac{l+1}{n}  \right )^{3/2}\right ) \right ] .
\end{split}
\end{equation}

The first term in Eq.~\ref{eq:QED} corresponds to vacuum polarization and the second term to the self energy. The latter gives rise to the anomalous magnetic moment of the electron. The electron's $g$-factor is $2+\alpha/\pi$ (to lowest order). This changes the electron's magnetic moment, leading to the second term in squared brackets in Eq.~\ref{eq:QED}, which is equivalent to accounting for the lowest-order correction of the electron's $g$-factor in Eqs.~\ref{eq:FShamiltonian} and ~\ref{eq:FSshift}.

In Eq.~\ref{eq:QED}, the self-energy term is typically two orders of magnitude higher than the vacuum polarization term, with the values of the circular state of interest being 0.59~Hz and 1.1~mHz, respectively, and values for the near-circular state being 0.31~Hz and 0.89~mHz. These corrections are small and lead to a small transition energy shift due to the Lamb shift as shown in Table~\ref{tab:budget}.

\subsection{Blackbody Shift}

Blackbody radiation has two effects on Rydberg atoms: the on-resonant portion affects their lifetime and the off-resonant portion can cause an energy shift \cite{Farley1981}. Since Rydberg-Rydberg transition frequencies are in the range of thermal blackbody radiation, thermal transitions lead to a lifetime reduction. For instance, the lifetime of the $n=50$ circular state is reduced from 30~ms at 0~K to 10~ms at 4~K. Since both states in the transition of interest (Fig.~\ref{fig:wavefunction}) are affected by this lifetime reduction, we expect a linewidth in the range of 30~Hz, which is sufficiently narrow for the current purpose.

The blackbody shift caused by the off-resonant portion is potentially of greater concern. The blackbody energy shift is given by
\begin{equation}
\label{eq:Ebbr}
E^{nl}_{\mathrm{BBR}} = \frac{e^2}{\hbar }\sum_{n',l'} \int_0^\infty \frac{|R^{n'l'}_{nl}|^2 |\mathbf{F}_{\textrm{b}}|^2 \Delta\omega}{2(\Delta\omega^2-\omega_{\textrm{b}}^2)} d\omega_{\textrm{b}},
\end{equation}
\noindent where $\omega_{\textrm{b}}$ is the angular frequency of the blackbody radiation, $|\mathbf{F}_{\textrm{b}}|^2$ is the blackbody field-amplitude squared per frequency unit, $R^{n'l'}_{nl}$ is the radial matrix element, and $\Delta\omega$ is the transition angular frequency difference between the final and initial states being considered. The field amplitude, $|\mathbf{F}_{\textrm{b}}|^2$, can be obtained using the spectral energy density form of the Planck radiation law
\begin{equation}
\left | \mathbf{F}_{\textrm{b}} \right |^2= \frac{2\hbar \omega_{\textrm{b}}^3}{\epsilon _0 \pi^2 c^3 ( e^{\hbar \omega_{\textrm{b}}/ k_{\textrm{B}} T}-1 )} ,
\end{equation}
\noindent where $k_{\textrm{B}}$ is Boltzmann's constant and $T$ is the temperature. Eq.~\ref{eq:Ebbr} has an implicit dependency on the states being considered since $\Delta\omega$ is defined by the transition in question.

Approximations for the limiting cases of Eq.~\ref{eq:Ebbr} are given in \cite{GallagherBook}. In our case, the transition frequency of interest is about 100 GHz, which is on the order of the peak of the radiation spectrum at 4~K. As a result, in order to calculate the blackbody shift, Eq.~\ref{eq:Ebbr} has to be explicitly evaluated. For treating parabolic states, Eq.~\ref{eq:states} can be used.

\subsection{Hyperfine Structure Correction}

The interaction of the nuclear magnetic moment with the magnetic field caused by the valence electron gives rise to the hyperfine-structure Hamiltonian \cite{Arimondo1977}
 \begin{equation}
\begin{split}
\hat{H}_{\mathrm{HFS}} &= \frac{\mu_0 g_{\textrm{I}} e^2}{4\pi m_{\textrm{e}} m_{\textrm{p}}} \left ( \frac{\hat{\mathbf{L}}\cdot \hat{\mathbf{I}}}{2r^3} -\frac{g_{\textrm{e}}}{4r^3}  \hat{\mathbf{S}}\cdot \hat{\mathbf{I}} \right. \\
&\quad \left. + \frac{g_{\textrm{e}}}{4r^3} 3(\hat{\mathbf{S}}\cdot \hat{\mathbf{r}})(\hat{\mathbf{I}}\cdot \hat{\mathbf{r}}) + \frac{g_{\textrm{e}}}{3} \frac{\delta(r)}{r^2} \hat{\mathbf{S}}\cdot\hat{\mathbf{I}} \right ),
\end{split}
\end{equation}
\noindent which acts on the space $\{ |n,n_{\textrm{1}},n_{\textrm{2}},m_{\textrm{l}},m_{\textrm{s}},m_{\textrm{i}} \rangle \}$, where $m_{\textrm{i}}$ is the nuclear magnetic quantum number. Above,  $\mu_0$ is the permeability of free space, $g_{\textrm{I}}$ is the $g$-factor of the nucleus, $m_{\textrm{p}}$ is the mass of the proton, $\hat{\mathbf{I}}$ is the nuclear spin operator and $\delta(r)$ is the Dirac delta function. The last term in the Hamiltonian is a contact term, where the energy depends on the wavefunction density at the position of the nucleus, which is zero for the states we consider in this experiment (Fig.~\ref{fig:wavefunction}).

Noting that the hyperfine structure is in the Paschen-Back regime and using first-order perturbation theory and the analytic expression given in \cite{BetheBook} for the $r^{-3}$ matrix elements, we obtain a hyperfine energy shift of
\begin{equation}
\begin{split}
E_{\mathrm{HFS}} &= m_{\textrm{i}}\sum_l  \frac{| C^{n_{\textrm{1}}n_{\textrm{2}}}_{lm_{\textrm{l}}} |^2}{a_0^3 n^3(l+1)(l+\frac{1}{2})l}   \\
&\quad  \times \frac{\mu_0 g_{\textrm{I}} \hbar^2  e^2}{8\pi m_{\textrm{e}} m_{\textrm{p}}}\left ( m_{\textrm{l}} - \frac{g_{\textrm{e}} m_{\textrm{s}}}{2} \right. \\
&\quad \left. \times \left [ 1-3\frac{2l^2+2l-2m_{\textrm{l}}^2-1}{(2l+3)(2l-1)} \right ]  \right ) ,
\end{split}
\end{equation}
\noindent which leads to a negligible energy shift (see Table~\ref{tab:budget}).

\subsection{Doppler Effect}

\begin{figure}[t]
\begin{center}
\includegraphics[width=3.4in]{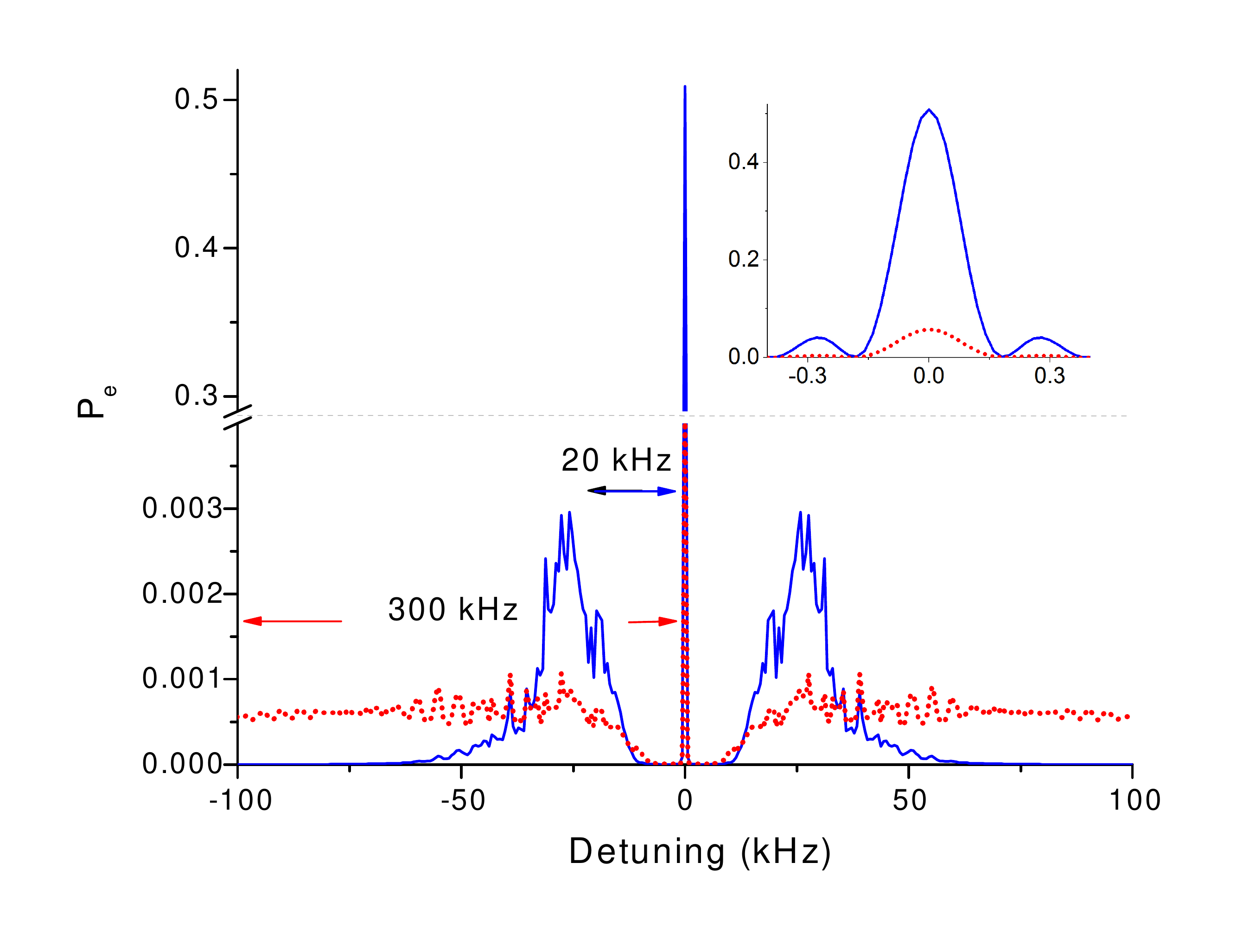}
\vspace{-0.3 in}
\caption{\label{fig:doppler}Simulation of the excited-state population, $P_e$, as a function of detuning for two temperature regimes: 100~$\mu$K (red dashed line) and 1~$\mu$K (blue solid line) for a potential depth of 35~kHz (motivated by Table~\ref{tab:shiftScales}) and an interaction time of 5~ms. The inset shows that the widths of the narrow features at the center are Fourier-limited. The arrows indicate the approximate half widths of the Doppler-broadened background signals.}
\vspace{-0.2 in}
\end{center}
\end{figure}

To drive transitions, one of the components of the optical lattice is amplitude-modulated (see Fig.~\ref{fig:potential}; optical carrier angular frequency $\omega$ and modulation angular frequency $\Omega$). Ponderomotive spectroscopy involves the inelastic scattering of two counter-propagating optical-lattice photons of angular-frequency difference $\Omega$ \cite{MooreTBA}, which is at the atomic transition frequency. In traditional Raman spectroscopy with counter-propagating beams, the Doppler shift between the drive frequency experienced by the atoms, $\Omega'$, and $\Omega$ would follow the expression
\begin{equation}
\label{eq:Doppler}
\Omega'- \Omega \approx - 2\omega \hat{\mathbf{k}} \cdot \frac{\mathbf{v}}{c}+\frac{\Omega}{2}\frac{v^2}{c^2},
\end{equation} 
\noindent where $\mathbf{v}$ is the center-of-mass velocity of an atom and the unit vector $\hat{\mathbf{k}}$ marks the direction of propagation of the beams. While the second-order Doppler effect is entirely negligible at MOT temperatures, the first-order Doppler effect would lead to a shift of about 20~kHz (for temperature of 1~$\mu$K). In ponderomotive spectroscopy Eq.~\ref{eq:Doppler} does not apply. While it approximately accounts for the overall widths of the spectra (indicated by the arrows in Fig.~\ref{fig:doppler}), it fails to describe the central, narrow peak observed in ponderomotive spectroscopy. There, the cooling and trapping of the atoms in a magic lattice \cite{MooreMagic}, combined with the fact that the phase of the Rabi frequency is constant within a potential well, allow us to achieve Doppler-free, Fourier-limited line-widths of the central peak (see Fig.~\ref{fig:doppler}).

In order to model the spectra, we employ a simulation program that treats the center-of-mass dynamics of the atoms (due to lattice-induced forces) classically and the internal, modulation-driven dynamics quantum-mechanically \cite{Moore2015, MooreMagic, Trajectories2010}. The effects of temperature on the population fraction that becomes excited into the upper state are shown in Fig.~\ref{fig:doppler} for a potential depth of 35~kHz in a one-dimensional lattice. The central, Fourier-limited features shown correspond to trapped atoms (no Doppler effect). As the temperature is lowered, the fraction of atoms trapped in the optical-lattice wells increases, leading to a corresponding increase of the area under the central peaks in Fig.~\ref{fig:doppler}. For a substantial number of atoms to be captured, molasses temperatures are required, as shown in Fig.~\ref{fig:doppler}. At 100~$\mu$K, only about 6$\%$ of the population is trapped, whereas 1~$\mu$K yields 51$\%$ trapped population. When temperatures are lowered, the full width of the Doppler-broadened background signal is $\approx 4v\omega/\pi c$, where the thermal velocity $v=\sqrt{k_{\textrm{B}}T/M_{\textrm{atom}}}$ (see Eq.~\ref{eq:Doppler}). The gaps between the central peak and the onset of the Doppler background reflect the fact that atoms within a range of velocities are trapped. The trapped atoms experience no Doppler shift and generate the Fourier-limited feature at the center of the spectrum. They essentially undergo recoil-free absorption within the lattice wells. A more complete discussion of this topic will be given in a future paper.

\section{Discussion}

\begin{table*}
\caption{\label{tab:budget} Transition frequency shifts, relative transition shifts and relative uncertainties for ground-based experiment under conditions suitable for a kinetic temperature of 1~$\mu$K. Improved micro-gravity-conditions shifts and uncertainties are shown in square brackets for a temperature of 10~nK. The second-order Stark and diamagnetic shifts are lowered under these conditions since the field values are determined based on the kinetic temperature of the atoms (see section~II.B). Here, we use $m_s=1/2$.}
\begin{tabular}{ llll}
 Source  & $\Delta \nu$  &   $\Delta \nu/\nu$  & $\delta \Delta \nu/\nu$ ($\times10^{-12}$) \\
\hline
Residual Lattice Shift            & 0 (3)			&0							& 32	          \\
 				   & [0 (0.1)] 			&	  						&  [1.0]	     \\
Dipolar Polarization Quantum Defect	   &120.1(3) Hz		&$1.283 \times 10^{-9}$				&  2.8  	    \\
2$^{nd}$ order Stark           &  -6.8 (1) Hz              	&$-7.3 \times10^{-11}$				& 1.5	        \\
		  		   & [-0.73 (1) mHz]	&       	     							& [$1.6 \times10^{-4}$]	\\
Diamagnetic		              & 0.94 (4) Hz  	           &$1.0 \times10^{-11}$		 		& 0.4           \\
 				   & [94 (4) $\mu$Hz]	&							&  [$4.0 \times10^{-5}$]     \\
Mass Correction	              & -605.08747 (3) kHz	&$6.4606271 \times10^{-6}$	  		& 0.3	           \\
Lamb Shift		              & -84.1 (5) mHz  		&$-8.98 \times10^{-13}$				& $5.0 \times 10^{-3}$         \\
Blackbody \footnote{at 4 K.} & 0.64 (6) mHz 		&$6.8 \times10^{-15}$				& $6.2 \times 10^{-4}$
        \\
Quadrupolar Polarization Quantum Defect  & 26 (5) $\mu$Hz 		& $2.8 \times10^{-10}$				& $6.0 \times 10^{-5}$
        \\
2$^{nd}$ order Doppler		   & 0.05 (7)nHz		  	&$5 \times10^{-22}$	          	 & $7.3 \times10^{-10}$	      	     \\
Fine Structure    		   &488.0332466612(5)Hz		&$5.210818188587 \times 10^{-9}$	 		& $1.6 \times 10^{-9}$             \\
Hyperfine Structure              &32.89402(7)  $\mu$Hz	&$3.512153 \times 10^{-16}$			& $7.2 \times 10^{-12}$          \\		
1$^{st}$ order Stark            & 0		  		&0							 & 0	          \\
 and Zeeman			   & 0		  		&0						          	 & 0	      	     \\
1$^{st}$ order Doppler		   & 0		  		&0						          	 & 0	      	     \\
\end{tabular}
\end{table*}

In Table~\ref{tab:budget}, we summarize the sources of frequency shifts and their respective relative uncertainties for the sample atomic transition $|51,0,0,50 \rangle \rightarrow |53,1,1,50 \rangle$ considered in this paper. These lead to an expected relative uncertainty in the proposed measurement of the Rydberg constant in the low $10^{-11}$ range. In the following, we discuss the leading sources of uncertainty and how these can be improved in order to attain a state-of-the-art uncertainty. In contrast with measurements performed with low-lying states of hydrogen (from which the best current uncertainty of $5.9 \times 10^{-12}$ is obtained), our measurement is independent of nuclear effects and therefore could contribute to solving the proton radius puzzle \cite{ScientificAmerican2014}.

\subsection{Main sources of uncertainty}

The main source of uncertainty on the proposed ground-based experiment is the residual lattice-induced shift. The resulting uncertainty is $3.2 \times 10^{-11}$; it is mainly due to laser-intensity fluctuations. This uncertainty value is obtained assuming a 1$\%$ lattice-intensity uncertainty (better stabilities are likely possible). The lattice shift results presented in Table~\ref{tab:budget} are obtained through simulations based on Eq.~\ref{eq:3DPOL} where we specify the two atomic states of interest, arbitrary laser-beam geometry, wavelengths and intensities of the beams that form the lattice. For these calculations, it is assumed that the alignment between the normal vectors of the lattice planes and the quantization axis is perfect. As suggested by Fig.~\ref{fig:potential}, conditions can be chosen such that one pair of lattice beams causes a transition shift with a different polarity than that due to another pair of lattice beams, such that the induced shifts can cancel each other out. To achieve this type of magic lattice, the intensity of one of the beams is adjusted until the calculated transition lattice shift reaches zero. In Table~\ref{tab:budget}, the corresponding uncertainties are generated by the assumption that the intensities can be controlled with a 1$\%$ relative uncertainty. 

Smaller uncertainties are possible through the use of shallower lattices and fewer laser beams. To achieve this, lower atomic temperatures are needed, for which other well-known cooling methods can be employed \cite{MetcalfBook}. Moreover, for a similar experiment in micro-gravity conditions, the depth of the lattice could be further decreased or even only used for driving transitions. This reduction of the lattice-induced uncertainty may be the only improvement necessary to achieve a competitive Rydberg-constant measurement. In Table~\ref{tab:budget}, lattice-shift estimates are presented for two cases of the lattice depth, one suitable for ground-based experiments and the other for micro-gravity experiments. Table~\ref{tab:budget} also shows that the lattice-induced shift represents the by far dominant source of systematic uncertainty and needs to be addressed first in any incremental improvement of the experiment.

The next-significant systematic shift arises from the dipole polarizability of the ionic core (see Table ~\ref{tab:budget}), which currently stands at $9.12(2)$ \cite{GallagherNew}, with experiments underway to improve the uncertainty in this value \cite{MooreTBA}. The uncertainty due to the quadrupolar energy level shift is negligible in the overall uncertainty budget, because of the $1/r^{6}$ dependence of that shift \cite{Gallaghernsnp, Gallaghernf}.

The quantum defect corrections are due to deviations from the hydrogenic $1/r$ potential. This applies to both the penetration \cite{GallagherBook} (zero for the case of circular states) and the polarization quantum defects. This issue could be avoided in the first place by using hydrogen instead of rubidium. However, experimental obstacles due to the large recoil energy of hydrogen and laser-cooling on the Lyman-$\alpha$ line are prohibitive at this time, leaving rubidium as an attractive option.

The second-order Stark shift leads to an uncertainty due the electric field not being precisely known. For the values displayed in Table ~\ref{tab:budget}, we assume an electric-field uncertainty of 1$\%$ of the field magnitudes provided in Table~\ref{tab:fields}. The resulting uncertainty is the third-largest in Table ~\ref{tab:budget}. It is seen in Table~\ref{tab:budget} that by performing the experiment under micro-gravity conditions the second-order Stark shift uncertainty can be decreased by four orders of magnitude (because the electric field can be dropped by two orders of magnitude) and hence the shift goes from border-line significant to insignificant.

\subsection{Other sources of uncertainty}

The following discussion shows that the remaining shifts listed in Table ~\ref{tab:budget} present negligible uncertainties at the current level of precision ($5.9 \times 10^{-12}$), but these are discussed here for completeness.

The uncertainty displayed for the second-order Zeeman shift in Table ~\ref{tab:budget} assumes the magnetic field is known to 2$\%$ of its value that is dictated by the kinetic temperature of the atoms (see Table~\ref{tab:fields}). This precision can be achieved by monitoring the atomic Larmor precession of cold-atom samples using the Faraday rotation technique.

The finite-mass correction, which accounts for the non-infinite mass of the nucleus, consists of a dominant first-order term and several higher-order terms. The first order can be considered by multiplying the Rydberg constant by a factor of $\mu/m_{\textrm{e}}$ ($\mu= m_{\textrm{e}}M/(m_{\textrm{e}}+M)$) \cite{BetheBook}, where $M$ is the Rb$^{+}$ mass. The correction is $-605.08747$~kHz, as shown in Table~\ref{tab:budget}. The mass correction introduces an insignificant uncertainty to our measurement since the mass of Rb$^+$ (84.911 245 324 a.u.) and that of the electron are well known (relative uncertainties of $4.4 \times10^{-8}$ \cite{Mount2010} and $1.2 \times10^{-8}$ \cite{CODATA2014}, respectively). The higher-order terms show up as factors of the form $(\mu/m_{\textrm{e}})^{\eta}$ in the fine structure ($\eta=1$), second-order Stark effect ($\eta=3$), diamagnetic shift ($\eta=1$), Lamb shift ($\eta=2$) and hyperfine structure ($\eta=1$) corrections. When the mass correction factor is considered for these, the shifts decrease by 3~mHz, 0.1~mHz, 6~$\mu$Hz, 1~$\mu$Hz, and 212~pHz, respectively. Since these differences are negligible, we do not carry out these corrections in the results shown in Table~\ref{tab:budget}. 

In contrast to measurements on low-lying states of hydrogen \cite{CODATA2002}, through the use of circular states, we obtain low QED corrections, since the valence electron has zero probability of being in the vicinity of the nucleus, and the size of the Rydberg function becomes very large. The main source of uncertainty in the Lamb shift is the Bethe logarithm (relative uncertainty of $3.3 \times 10^{-8}$ and $1.0 \times 10^{-7}$ or less for the circular and near-circular states, respectively \cite{Erickson1977}), leading to a negligible uncertainty for our measurement.

The blackbody shift is lowered three orders of magnitude by placing our system in thermal contact with liquid helium, which has a temperature of 4~K. Even so, at 300~K the blackbody radiation shift for the transition of interest is just $-21$~mHz, making this shift negligible for a wide range of typical experimental temperatures. The results of numerical calculations shown in Table~\ref{tab:budget} follow a similar procedure to that presented in \cite{Farley1981}. However, we consider only bound states up to about $n$=300. This truncation of the basis set does not affect the results significantly. By leaving out the last 250 states in the calculation, at worst, the calculated shift for the individual states (about 2.4~kHz at 300~K and 0.42~Hz at 4~K) changes only by about 0.1~mHz at a temperature of 300~K and by nano-Hertz at a temperature of 4~K. This leads us to the conclusion that it is not necessary to include the continuum states in our calculations, which is also reaffirmed in \cite{Farley1981}. We treat the near-circular state as a sum of spherical states multiplied by their respective 3J symbols squared. The radiation field is taken to be isotropic inside our spectroscopy enclosure, since at the frequencies considered, the cavity density of states approaches that of free space. With this treatment, we obtain results comparable to those obtained in \cite{Farley1981} for temperatures of 300~K. The radial matrix elements used in the calculations are correct to four significant figures. The main source of uncertainty for the shift  presented in Table ~\ref{tab:budget} is dictated by how well we know the temperature inside the spectroscopy enclosure. When calculating the blackbody shift uncertainty, it is assumed that the temperature is known to $\pm 0.5$~K.

The main disturbance caused by the blackbody radiation is the broadening it induces on the spectral line due to thermally-induced decays and excitations. A precise determination of the location of our measured frequency is essential in obtaining a successful measurement of $R_{\infty}$. Assuming a good signal-to-noise ratio, we expect to determine the line center to within $1/100$ of the line-width. At a radiation temperature of 4~K, and for the states considered in this paper, the corresponding statistical uncertainty is 0.3~Hz, corresponding to a relative uncertainty of $3 \times10^{-12}$.

Despite producing a relatively large shift, the fine-structure correction can be calculated very accurately since the fine-structure constant is well known (relative uncertainty of $2.3 \times10^{-10}$ \cite{CODATA2014}). As a result, the relative uncertainty introduced by the fine-structure shift is only $1.6 \times10^{-21}$.

Even though the hyperfine-structure shift in itself is negligible, its uncertainty is nevertheless estimated. The main sources of this uncertainty are the electron mass, proton mass and Planck's constant (all of them with relative uncertainties of $1.2 \times 10^{-8}$) and the electron charge (relative uncertainty of $6.1 \times 10^{-9}$) \cite{CODATA2014}. The $g$-factors for the nucleus (— 0.000 293 640 0) and the electron (−2.002 319 304 361 53) are also well known (relative uncertainties of $6.0 \times 10^{-10}$ \cite{Arimondo1977} and $2.6 \times 10^{-13}$ \cite{CODATA2014}, respectively).

Effective trapping of atoms in the lattice leads to zero first-order and negligible second-order Doppler effects (Fig.~\ref{fig:doppler}). As discussed in section~III.B, the first-order Stark and Zeeman shifts are zero for our transition. This is possible by choosing the lower and upper state so that either the shift on each individual state is zero (first-order Stark shift) or both states experience the same shift (first-order Zeeman shift).

\section{Conclusion}

We have discussed an experimental method to help solve the proton radius puzzle using a cold-atom-based measurement of the Rydberg constant ($R_{\infty}$), which utilizes (near-)circular Rydberg states and is free of QED shifts and sensitivity to nuclear charge overlap. Previous efforts to measure $R_{\infty}$ with Rydberg atoms have encountered several experimental challenges which are addressed in this proposed measurement. The first-order Zeeman and Stark shifts are both zero, owing to appropriate selection of parabolic atomic states involved in the transition. By applying cooling and trapping techniques, the interaction times are increased, leading to a reduction of the Fourier width. By using a new method of spectroscopy in modulated optical lattices, the Doppler broadening is eliminated. An implementation of the proposed experiment at atomic temperatures of 1~$\mu$K is projected to yield a Rydberg constant value with an uncertainty in the low $10^{-11}$ range, limited almost exclusively by lattice-trap-induced shifts. Since the proposed experiment differs from spectroscopy on low-lying atomic states, in that it is entirely insensitive to the proton radius, this level of precision would be sufficient to make a statement about the proton radius puzzle. The trap-induced shifts could be very well addressed by performing the experiment under micro-gravity conditions, which could lead to an uncertainty in the $\approx 3 \times 10^{-12}$ range, almost a factor of two improvement over the current uncertainty.

\section{Acknowledgments}

A.R. acknowledges support from the National Science Foundation Graduate Research Fellowship under Grant No. DGE 1256260. K.M. acknowledges support from the University of Michigan Rackham Pre-doctoral Fellowship. This work was supported by NSF Grant No. PHY1506093, and NASA Grant No.NNH13ZTT002N NRA.


\bibstyle{apsrev4-1}

\providecommand{\noopsort}[1]{}\providecommand{\singleletter}[1]{#1}%

\end{document}